On the fundamentals of Richtmyer-Meshkov dynamics with variable acceleration

Aklant K. Bhowmick (1); Desmond L. Hill (2); Miccal Matthews (2); Snezhana I. Abarzhi* (2)

Carnegie Mellon University, USA (1); University of Western Australia, AUS (2)

*corresponding author snezhana.abarzhi@gmail.com

Richtmyer-Meshkov instability (RMI) plays important role in nature and technology, from supernovae and fusion to scramjets and nano-fabrication. Canonical Richtmyer-Meshkov instability is induced by a steady shock and impulsive acceleration, whereas in realistic environments the acceleration is usually variable. This work focuses on RMI induced by acceleration with a power-law time-dependence, and applies group theory to solve the classical problem. For early-time dynamics, we find the dependence of RMI growth-rate on the initial conditions and show it is free from the acceleration parameters. For late time dynamics, we find a continuous family of regular asymptotic solutions, including their curvature, velocity, Fourier amplitudes, and interfacial shear, and we study the solutions stability. For each of the solutions, the interface dynamics is directly linked to the interfacial shear, and the non-equilibrium velocity field has intense fluid motion near the interface and effectively no motion in the bulk. The quasi-invariance of the fastest stable solution suggests that nonlinear coherent dynamics in RMI is characterized by two macroscopic length-scales - the wavelength and the amplitude, in excellent agreement with observations. We elaborate new theory benchmarks for experiments and simulations, and put forward a hypothesis on the role of viscous effects in interfacial nonlinear RMI.

Rayleigh-Taylor instability (RTI) develops at the fluid interface when fluids of different densities are accelerated against their density gradients; Richtmyer-Meshkov instability (RMI) develops when the acceleration is induced by a shock and is impulsive [1-4]. Intense interfacial Rayleigh-Taylor (RT) / Richtmyer-Meshkov (RM) mixing of the fluids ensues with time [5,6]. RTI/RMI and RT/RM mixing play important role in a broad range of processes in nature and technology, including stellar evolution, plasma fusion, and fossil fuel industry [7-13]. In this work we study the long-standing problem of RMI with variable acceleration [14]. We employ group theory to solve the boundary value problem for the early- and late-time RM evolution [15], directly link RM dynamics to the interfacial shear, identify its invariance properties, and reveal the interfacial and multi-scale character of RM dynamics. Our theory finds substantial differences between RM and RT dynamics with variable accelerations [14], excellently agrees with existing observations, and elaborates new diagnostic benchmarks for experiments and simulations.

RMI with variable acceleration commonly occur in fluids, plasmas, materials [7-14]: RMI leads to appearance of light-years-long structures in clouds of molecular hydrogen, influences formation of hot spot in inertial confinement fusion, controls combustion processes in scramjets, and drives materials' transformation under impact in nano-fabrication. In these vastly different physical conditions, RM flows have similar qualitative features of their evolution. The post-shock RM dynamics is a superposition of two motions, which are the background motion of the fluid bulk and the growth of the interface perturbations [1,2,6,16-19]. In the background motion, both fluids and their interface move as whole unit in the direction of transmitted shock; this motion occurs even for an ideally planar interface and is super-sonic for strong shocks. The growth of the interface perturbations is due to impulsive acceleration by the shock; it develops only when the flow fields are perturbed; its growth rate is sub-sonic and the associated motion is incompressible [1,2,6,16-23]. RM unstable interface is transformed to a composition of small-scale shear-driven vortical structures and a large-scale coherent structure of bubbles and spikes, where the bubble (spike) is a portion of the light (heavy) fluid penetrating the heavy (light) fluid. Small-scale non-uniform structures appear also in the bulk, including hot and cold spots, high and low pressure regions, cumulative jets, checker-board velocity patterns [5,6,16-20]. Self-similar RM mixing develops, and energy supplied initially by the shock gradually dissipates [5,6,16-23]. While RM and RT flows are similar in many regards, there are also important distinctions [3-6,15,20]. RTI is driven by the acceleration, and it starts to develop when the flow fields and/or the interface are slightly perturbed near the equilibrium state. RT flows are free from the background motion. Interfacial RT mixing induced by constant acceleration accelerates, whereas interfacial RM mixing induced by impulsive acceleration decelerates [1-6,14-23].

RMI/RTI and RM/RT mixing are a challenge to study in theory, experiments and simulations [14-34]. In theory we have to develop new approaches for non-equilibrium multi-scale RM/RT dynamics, identify properties of asymptotic solutions, and capture symmetries of RM/RT flows [5,15,25-30]. Experiments have to meet tight requirements for the flow implementation, diagnostics and control [2,6,20-24]. Simulations have to employ highly accurate numerical methods and massive computations for capturing shocks, tracking interfaces, and accurately modeling small-scale processes over a large span of scales [16-19,30-34]. Besides, substantial span of temporal and spatial scales is required for bias-free interpretation of experimental and numerical data describing RM/RT evolution [14-34]. Significant success has been recently achieved in the understanding of RMI and RTI as well as RM and RT mixing [5,15,25]. Particularly, group theory approach has found a multi-scale character of nonlinear RMI and RTI and an order in RT mixing with constant acceleration, thus explaining the observations [5,15,25].

In realistic environments, RT and RM flows are usually driven by variable acceleration [8-14]. For instance, in supernova blast, inertial confinement fusion and nano-fabrication, accelerations are induced by variable strong shocks [8-14,35,36]. Only limited information is currently available on RM and RT dynamics under these conditions [14]. An important case is acceleration with a power-law time-dependence, since power-law functions may lead to new scaling properties of the dynamics and be used to adjust the acceleration's time-dependence in applications [8-14,35,36]. For accelerations with power-law time-dependence, early-time and late-time dynamics can be RM or RT type depending on the exponent of the acceleration power-law [14]. Particularly, the interfacial dynamics is driven by the acceleration and is RT-type for exponent values larger than -2, and driven by the initial growth-rate and is RM-type otherwise [14].

In this work we study the long-standing problem of RMI with variable acceleration for a three-dimensional spatially extended periodic flow. We apply group theory to solve the boundary value problem involving boundary conditions at the interface and the outside boundaries and the initial value problem [5,14,15,25]. For early-time dynamics, we find the dependence of RMI growth-rate on the initial conditions and show it is free from the acceleration parameters. For late-time dynamics, we directly link the interface dynamics to the interfacial shear, find a continuous family of regular asymptotic solutions, and study the solutions stability. For each of the family solutions, the perturbed velocity field has intense fluid motion near the interface and effectively no motion in the bulk. We identify parameters of the critical, Atwood, Taylor and flat bubbles, including their curvature, velocity, Fourier amplitudes, and interfacial shear. The Atwood bubble – the fastest stable solution – has a quasi-invariance property suggesting that nonlinear coherent RM dynamics is set by the interplay of two macroscopic length-scales - the wavelength and the amplitude. We discuss connections of RM dynamics to the theory of plug flow in ideal fluids, and put forward a hypothesis on the role of viscous effects in the interfacial RMI. Our theory excellently agrees with existing observations, and elaborates new benchmarks for observations [6,8-14,20-23].

RM dynamics of ideal fluids is governed by the conservation of mass, momentum and energy

$$\partial\rho/\partial t + \partial\rho v_i/\partial x_i = 0,\ \partial\rho v_i/\partial t + \sum_{j=1}^{3}\partial\rho v_i v_j/\partial x_j + \partial P/\partial x_i = 0,\ \partial E/\partial t + \partial(E+P)v_i/\partial x_i = 0 \quad (1a)$$

with spatial coordinates $(x_1,x_2,x_3)=(x,y,z)$; time $t$; fields of density, velocity, pressure and energy $(\rho,\mathbf{v},P,E)$, with $E=\rho(e+\mathbf{v}^2/2)$ and $e$ being specific internal energy [14,35]. These equations are augmented with boundary conditions, including boundary conditions at the interface and at the outside boundaries of the domain, and with initial conditions prescribing initial perturbations of the flow fields in the bulk and at the interface [5,14,15,25-29,35].

To the leading order (i.e., for unperturbed interface and flow fields), the boundary conditions define the post-shock flow fields, including densities of the fluids, velocities of the transmitted and reflected shocks, and the velocity of the background motion of the fluids and the interface [16-19,28, 29].

In the frame of reference moving in the $z$ direction with the velocity of the background motion, boundary conditions at the interface and at the outside boundaries are

$$[\mathbf{v}\cdot\mathbf{n}]=0, [P]=0, [\mathbf{v}\cdot\mathbf{\tau}]= arbitrary, [W]= arbitrary, \mathbf{v}|_{z\to+\infty}=0, \mathbf{v}|_{z\to-\infty}=0 \quad (1b)$$

where $[...]$ denotes the jump of functions across the interface; $\mathbf{n}(\mathbf{\tau})$ are the normal and tangential unit vectors of the interface with $\mathbf{n}=\nabla\theta/|\nabla\theta|, (\mathbf{n}\cdot\mathbf{\tau})=0$; $\theta=\theta(x,y,z,t)$ is a local scalar function, with $\theta=0$ at the interface and $\theta>0$ ($\theta<0$) in the bulk of the heavy (light) fluid marked hereafter with subscript $h(l)$. Specific enthalpy is $W=e+P/\rho$. Density jump at the interface is parameterized by the Atwood number $A=(\rho_h-\rho_l)/(\rho_h+\rho_l)$ with $0<A\leq 1$ and $A\to 1^-(0^+)$ as $(\rho_l/\rho_h)\to 0^+(1^-)$ [1-6].

Initial perturbations cause the instability to develop and non-equilibrium RM flow to occur [5,6,14-20]. The flow is periodic in the plane $(x,y)$ normal to acceleration $\mathbf{g}, |\mathbf{g}|=g$., which is directed from the heavy to light fluid, $\mathbf{g}=(0,0,-g)$ along the $z$ axis and adds to pressure term $\rho g z$. The acceleration is a power-law function of time $g=Gt^a$, where $a$ is exponent, $a\in(-\infty,+\infty)$, $G$ is pre-factor, $G>0$, and dimensions are $\dim a=1$, $\dim G=m/s^{a+2}$. The acceleration can be due to an external force and/or be an effective acceleration due to variable velocity of the background motions induced by, e.g., variable shocks [14,20-36].

Initial conditions include initial perturbations of the interface and the flow fields with wavelength $\lambda$ and initial growth-rate $v_0$ at initial time $t_0$ [5,6,15,25]. For given $\lambda$ and $v_0$ for $a\neq -2$, the dynamics has two time-scales, $\tau_G\sim(\lambda/G)^{1/(a+2)}$ and $\tau_0\sim(\lambda/v_0)$; at $a=-2$ the time-scale is $\tau_0\sim(\lambda/v_0)$, and value $(G/\lambda)$ parameterizes the acceleration strength [14,29]. We presume that initial time is $t_0>\{\tau_G,\tau_0\}$. For $a>-2$ the time-scales are $(\tau_G/\tau_0)\ll 1$, and the fastest process is set by the acceleration; this is the acceleration driven Rayleigh-Taylor type dynamics. For $a<-2$ the time-scales are $(\tau_0/\tau_G)\ll 1$; the fastest process is set by the initial growth-rate; this is the initial growth-rate driven Richtmyer-Meshkov type dynamics. At $a=-2$ a transition occurs from RT to RM type dynamics with varying parameter $(G/\lambda)$. This work is focused on the initial growth-rate driven RM dynamics with $a<-2$. Time is $t>t_0>0, t_0\gg\tau$. The cases of RT-type dynamics for $a>-2$ and RT-RM transition at $a=-2$ are considered elsewhere [14,29].

We employ group theory to solve the problem of RMI with variable acceleration [5,14,15]. This approach solves the nonlinear boundary value problem and initial value problems with account for the non-local and singular character of RM dynamics, employs canonical forms of Fourier series and spatial expansions, and strictly obeys the conservation laws. See for details [5,14,15]. By using techniques of theory of space discrete groups (also known as Fedorov or Schoenflies groups), we first identify the groups which enable structurally stable dynamics, such as groups of hexagon p6mm, square p4mm, rectangle p2mm in 3D; group pm11 in 2D. We next apply irreducible representations of a relevant group to expand flow fields as Fourier series, and further make spatial expansions in a vicinity of a regular point at the interface. Governing equations are reduced to a dynamical system in terms of surface variables and moments. We solve the system in various asymptotic regimes [5,14,15,25,27].

We focus on the large-scale coherent dynamics with scales $\sim \lambda$ and presume that shear-driven interfacial vortical structures are small with scales $<< \lambda$ [5,14,15,25,27]. For convenience, we further transfer to the frame of reference moving with velocity $v(t)$ in the $z$-direction, where $v(t) = \partial z_0 / \partial t$ and $z_0$ are the velocity and position of the bubble (spike) in laboratory reference frame. For large-scale coherent structure fluid motion is potential. For symmetry group p6mm, the velocity is $\mathbf{v}_{h(l)} = \nabla \Phi_{h(l)}$, with $\Phi_h(\mathbf{r}, z, t) = \sum_{m=0}^{\infty} \Phi_m(t) \left( z + [\exp(-mkz)/(3mk)] \sum_{i=1}^{3} \cos(m\mathbf{k}_i \mathbf{r}) \right) + f_h(t) + cross\ terms$ and $\Phi_h \to \Phi_l$ upon $h \to l$, $\Phi_m \to \widetilde{\Phi}_m$, $z \to -z$. Here $\mathbf{r} = (x, y)$, $\mathbf{k}_i$ are the vectors of the reciprocal lattice with $\mathbf{k}_i \mathbf{a}_j = 2\pi \delta_{ij}$, $\mathbf{a}_i$ are the spatial periods in the $(x, y)$ plane, $|\mathbf{a}_i| = \lambda$, $k = |\mathbf{k}_i| = 4\pi/(\lambda\sqrt{3})$, $i, j = 1, 2, 3$, $\Phi_m (\widetilde{\Phi}_m)$ are the Fourier amplitudes of the heavy (light) fluid, $f_{h(l)}$ are time-dependent functions, and $m$ is integer. Fluid interface is $\theta = -z + z^*(x, y, t)$ with $z^* = \sum_{N=1}^{\infty} \zeta_N(t) \mathbf{r}^{2N} + cross\ terms$, where $\zeta_N$ are surface variables, $\zeta_1 = \zeta$ is the principal curvature at the bubble (spike) tip, $N$ is approximation order [5,15,25,27].

To the first order $N = 1$, the dynamical system is [5,15,25,27,38]

$$\rho_h (\dot{\zeta} - 2\zeta M_1 - M_2/4) = 0,\ \rho_l (\dot{\zeta} - 2\zeta \widetilde{M}_1 + \widetilde{M}_2/4) = 0,\ M_1 - \widetilde{M}_1 = any,\quad (2)$$

$$\rho_h (\dot{M}_1/4 + \zeta \dot{M}_0 - M_1^2/8 + \zeta g) = \rho_l (\dot{\widetilde{M}}_1/4 - \zeta \dot{\widetilde{M}}_0 - \widetilde{M}_1^2/8 + \zeta g),\ M_0 = -\widetilde{M}_0 = -v$$

where $M(\widetilde{M})$ are the heavy (light) fluid moments, $M_n(\widetilde{M}_n) = \sum_{m=0}^{\infty} \Phi_n(\widetilde{\Phi}_n) k^n m^n + cross\ terms$, each of which is an infinite sum of weighted Fourier amplitudes, and the interface is $z^* = \zeta(x^2 + y^2)$.

With length-scale $k^{-1}$ and time scale $\tau = \tau_0 = (kv_0)^{-1}$, for early-time dynamics, $(t - t_0) << \tau$, only first order harmonics are retained in the expressions for moments $M_n(\widetilde{M}_n) = k^n \Phi_1(\widetilde{\Phi}_1)$, $n = 0, 1, 2$.

For a nearly flat interface the solution is $(-\zeta/k) = (2A)^{-1}\ln(C_2(t/\tau) + C_1)$, $v = (4/k)d(-\zeta/k)/dt$ where $C_{1(2)}$ are integration constants defined by initial conditions $\zeta(t_0), v(t_0), |\zeta(t_0)/k|, |v(t_0)\tau k| \ll 1$. For $t \sim t_0$, $\zeta - \zeta(t_0) = -(k/4)[kv(t_0)(t-t_0)]$, $v - v(t_0) = -(A/2)v_0[(t-t_0)/\tau]$ suggesting that positions of bubbles (spikes) with $\zeta \leq 0, v \geq 0$ ($\zeta \geq 0, v \leq 0$) are defined by the initial velocity field, with bubbles (spikes) formed for $v(t_0)/v_0 > 0$ ($< 0$). For $a < -2$, with $\tau = \tau_0$ and $\tau_0 \ll \tau_G$, the instability growth-rate is independent of the acceleration parameters, since the contributions of acceleration-induced terms to early-time dynamics are negligible, $(-\zeta/k) \sim \sqrt{t/\tau_G}\, I_{\pm 1/2s}(\sqrt{A}\,(t/\tau_G)^s/s)$, $v = (4/k)d(-\zeta/k)/dt$, where $I_p$ is the modified Bessel function of the $p$th order and $s = (a+2)/2$ [14,38,39].

At late times, spikes are singular (for $0 < A < 1$ the singularity is finite-time) and bubbles are regular [5,14,15,25,27]. For $t \gg \tau$ regular asymptotic solutions depend on time as $v, M, \widetilde{M} \sim t^{-1}$ and $\zeta \sim k$, and higher order harmonics are retained in the expressions for moments. Group theory is further applied to solve the closure problem, find the family of regular asymptotic solutions, and identify properties of nonlinear RMI [5,14,15,25,27]. Note that for nonlinear dynamics the acceleration leads to time-dependence of regular asymptotic solutions $v, M, \widetilde{M} \sim (t/\tau_G)^{a/2}$; for $a < -2$, $\tau = \tau_0 \ll \tau_G$, these contributions are negligible when compared to $v, M, \widetilde{M} \sim t^{-1}$ induced by the initial growth-rate [14].

For the family of nonlinear regular asymptotic solutions, the bubble velocity $v \geq 0$ depends on its curvature $\zeta, \zeta < 0$, at $N = 1$ as

$$v = (kt)^{-1}(3 - 2A(\zeta/k)(-5 + 64(\zeta/k)^2))(9 - 64(\zeta/k)^2)(-48(\zeta/k) + A(9 + 64(\zeta/k)^2))^{-1} \quad (3a)$$

The function domain is $\zeta \in (\zeta_{cr}, 0)$, $\zeta_{cr} = -(3/8)k$, and the range is $v \in (v_{min}, v_{max})$, achieving value $v_{min} = 0$ at $\zeta = \zeta_{cr}$ and value $v = v_{max}$ at $\zeta = 0$. The velocity of RM bubbles is a monotone increasing function on curvature, $(\partial v/\partial \zeta) > 0$ for $\zeta \in (\zeta_{cr}, 0)$. The fastest solution corresponds to a flat bubble with $v_{max} = 3/(Akt)$, $\zeta_{max} = 0$. The multiplicity of nonlinear solutions is due to the singular and non-local character of RM dynamics. The number of the family parameters is set by the flow symmetry. For group p6mm the dynamics is highly isotropic, $z^* \sim \zeta(x^2 + y^2)$, and interface morphology is captured by principal curvature $\zeta$, Figure 1 [5,14,15,25,27,38,39]. At $N = 1$, the Fourier amplitudes are

$\Phi_1 = -4v(1 + 4(\zeta/k))/(3 + 8(\zeta/k))$, $\Phi_2 = (1 + 8(\zeta/k))/(3 + 8(\zeta/k))$, $\widetilde{\Phi}_1 = 4v(1 - 4(\zeta/k))/(3 - 8(\zeta/k))$ and $\widetilde{\Phi}_2/v = -(1 - 8(\zeta/k))/(3 - 8(\zeta/k))$.

In higher orders, $N>1$, solutions can be found similarly. For $N>1$ solutions exist and converge with increase in $N$. For $\zeta \in (\zeta_{cr},0)$ the lowest-order amplitudes $|\Phi_1|, |\tilde{\Phi}_1|$ are dominant, and the values of $|\Phi_m|, |\tilde{\Phi}_m|$ decay with increase of $m$. For $\zeta \sim \zeta_{cr}$ the convergence no longer holds [5,15].

The multiplicity of nonlinear regular asymptotic solutions is due to the presence of shear at the interface, Figure 1. We define shear $\Gamma$ as the spatial derivative of the jump of tangential velocity at the interface, $\Gamma = \Gamma_{x(y)}$, with $\Gamma_{x(y)} = \partial [v_{x(y)}]/\partial x(y)$, and find that $\Gamma = -M_1 + \tilde{M}_1$ near the bubble tip. Shear $\Gamma$ depends on the bubble curvature $\zeta$ as

$$\Gamma = t^{-1} 12 \big(3 - 2A(\zeta/k)\big(-5 + 64(\zeta/k)^2\big)\big)\big(-48(\zeta/k) + A\big(9 + 64(\zeta/k)^2\big)\big)^{-1} \quad (3b)$$

with $v = \Gamma k^{-1}\big((9 - 64(\zeta/k)^2)/12\big)$. For $\zeta \in (\zeta_{cr},0)$, shear $\Gamma$ achieves maximum value $\Gamma_{max}$ at $\zeta = \zeta_{max} = 0$, and minimum value $\Gamma_{min}$ at $\zeta = \hat{\zeta}_{min}$. For $A^* < A \le 1$, value $\hat{\zeta}_{min}$ is $\zeta_{cr} \le \hat{\zeta}_{min} < 0$ with $(\hat{\zeta}_{min}/k) = -(-3+\sqrt{22})/8$ at $A=1$ and $\hat{\zeta}_{min} = -(3/8)k$ at $A = A^* = 2/9$. For $0 < A < A^*$ value is $\hat{\zeta}_{min} = \zeta_{cr}$. To conveniently describe the solutions for $0 < A < 1$ we define the minimum value $\zeta_{min}$ as $\zeta_{min} = \hat{\zeta}_{min}$ for $A^* < A \le 1$ and $\zeta_{min} = \zeta_{cr}$ for $0 < A < A^*$. For $\zeta \in (\zeta_{min},0)$ the bubble velocity $v$ is the 1-1 function on interfacial shear $\Gamma$, $\Gamma \in (\Gamma_{min}, \Gamma_{max})$, and $v/\Gamma = (k/12)(9 + 64(\zeta/k)^2)$. We present elsewhere cumbersome function $v(\Gamma)$.

Our theory links with one another the interface morphology, shear and velocity in nonlinear RMI, and resolves the long-standing problem of multiplicity of asymptotic nonlinear solutions, Figure 1 [14,15,37,38]. Our theory further provides the value of the growth-rate of shear-driven Kelvin-Helmholtz instability (KHI) as $\omega_{KHI} \sim \Gamma$. Since $\Gamma \sim t^{-1}$ this suggests that the development of KHI is suppressed in a vicinity of RM bubbles with variable acceleration, in excellent agreement with experiments [6,14-20].

Consider some special solutions in the family. We call these solutions the critical bubble, the Taylor bubble, the Atwood bubble, and the flat bubble, Figure 1. For the critical bubble with curvature $\zeta_{cr} = -(3/8)k$, the solution is $v_{cr} = 0, \Gamma_{cr} = 2t^{-1}$ and $v_{cr}k = 0 \cdot \Gamma_{cr}$, and is independent of the Atwood number. The family has a solution which we call the 'Taylor bubble' since its curvature is the same as in Ref.[4] except for the difference in the wavevector value. For the Taylor bubble the solution is $\zeta_T = -(1/8)k$, $v_T(kt) = 4(3-A)/(3+5A)$, $\Gamma_T t = 6(3-A)/(3+5A)$ with $v_T k = (2/3)\Gamma_T$. At $N=1$ for the Taylor's bubble the Fourier harmonics $(\Phi_2)_T = 0$ is zero; for $N > 1$ this property does not hold. The Taylor's bubble velocity is close to that of the so-called Layzer-type bubble, with which experiments and simulations tend to compare well [40]. We call the fastest solution $(\zeta,v)_{max}$ the 'Atwood bubble' to emphasize its dependence on the Atwood number, $(\zeta,v)_A = (\zeta,v)_{max}$. For the Atwood bubble $(\zeta,v,\Gamma)_A = (\zeta,v,\Gamma)_{max}$ with $v_A k = (3/4)\Gamma_A$ and $\Gamma_A t = 4/A$. For the flat bubble, $\zeta_f = 0$, the solution is the same as for the Atwood bubble, $(\zeta,v,\Gamma)_f = (\zeta,v,\Gamma)_A$. For the Taylor and Atwood bubbles velocities and shears of are decreasing functions of the Atwood number, in agreement with available observations [6,20-22,29,34].

Analysis of asymptotic stability of solutions, with small departures $\sim \Delta(\beta,(t/\tau))$ and (un)stable solutions for $(\mathrm{Re}[\beta]>0)\,\mathrm{Re}[\beta]<0$, finds that at $N=1$ the family solutions are stable; the fastest stable solution $(\zeta,v,\Gamma) \sim (\zeta,v,\Gamma)_A$ is the physically significant solution, Figure 1. This solution has the (quasi) invariant value $(4/3) t\, v_A^2 \big/ \left| (dv/d\zeta)_{\zeta=\zeta_A} \right| = \left(1 + (5/2)(A/2)^2\right)^{-1} \approx 1$.

This (quasi) invariance implies that nonlinear RM dynamics is multi-scale, with two macroscopic length scales contributing – the wavelength and the amplitude [5,15,25]. The multi-scale character of the dynamics can be understood by viewing RM coherent structure as a standing wave with the growing amplitude [35]. The multi-scale character of nonlinear RMI is consistent with the existence of amplitude scale in early-time RMI, at which the maximum initial growth-rate of RMI is achieved [18]. Our theory further finds that non-equilibrium RM dynamics with variable acceleration is essentially interfacial: It has intense fluid motion in a vicinity of the interface, effectively no motion away from the interface and vortical structures produced by shear at the interface, Figure 2. This velocity pattern is observed in experiments and simulations, in excellent agreement with our results [16-24].

Similarly, solutions can be found for other symmetries in 3D and 2D flows. 3D flows tend to conserve isotropy in the plane. 3D highly symmetric dynamics is universal, except for the difference in the wavevector value. For 3D low symmetric dynamics with group p2mm, there is a two-parameter family of regular asymptotic solutions, and only nearly isotropic bubbles are stable. Dimensional 3D-2D crossover is discontinuous [5,15,25].

Our theory finds that in RMI with variable acceleration, nonlinear bubbles decelerate and flatten. Flattening and deceleration of RM bubbles is observed in experiments and simulations, in excellent agreement with our results [5,6,15-20]. According to our theory, in nonlinear RMI with variable acceleration flattened bubbles move quicker and decelerate stronger when compared to curved bubbles: Since bubble velocity decays with time as $\sim C/kt$, the deceleration is $\sim -C/kt^2$. According to our results, the Atwood and Taylor (as well as Layer-type) RM bubbles move quicker and have larger interfacial shear for fluids with similar densities rather than for fluids with very different densities [6,15,20,29,34,39]. This result has clear interpretation: for fluids with similar densities shear-driven interfacial vortical structures are more intense leading to stronger energy dissipation, stronger deceleration and larger bubble velocity when compared to the case of fluids with very different densities.

Our analysis is focused on large-scale dynamics presuming that interfacial vortical structures are small. This assumption is applicable for fluids very different densities and with a finite density ratio. For fluids with very similar densities $A \to 0^+$ other approaches should be employed [40]. While for fluids with very similar densities our theory is no longer applicable, a singular character of the velocity of the fastest stable Atwood bubble for $A \to 0^+$ indicates that for $A \to 0^+$ and $t/\tau \to \infty$ the bubble velocity may decay quicker than inverse time [5,6,15,20,25,38].

Singular character of the interfacial shear for the Atwood bubble at $A \to 0^+$ is consistent with the theory of plug flow in an ideal fluid, in which an infinitely large shear occurs in a vicinity of a flat immovable boundary [35,41]. The singularity of the plug flow is associated with absence of a boundary layer in ideal fluid and is akin to the d'Alembert's paradox [41]. Just as the d'Alembert's paradox is resolved by Prandtl's discovery that flow at high Reynolds number is a singular perturbation problem [41], an attempt to model RM dynamics with variable acceleration at any density ratio may require us to accurately account for the viscous effects. We address the testing of this hypothesis to the future. We note that the importance of viscous effects in RM dynamics is qualitatively consistent with the recently found sub-diffusive (i.e., slower than diffusion or dissipation) character of self-similar RM mixing with variable acceleration, in which larger velocities correspond to smaller length scales [14].

According to our results, for variable acceleration with $a < -2$, RM dynamics depends on the initial conditions and is independent of the acceleration. Hence, one can scrupulously study the effect of initial conditions on RM dynamics by analyzing properties of unstable interface for various accelerations [8-14]. Note that accurate quantification of nonlinear RMI in observations may be a challenge, since the interface velocity is usually ~0.1% of the largest velocity scale in the post-shock fluid system, and since the interface velocity is a power-law function of time, which requires substantial span of temporal and spatial scales for accurate diagnostics [15-20]. In addition to the interface velocity, we elaborate theory benchmarks which have not been discussed before. These are the fields of velocity and pressure, interface morphology and bubble curvature, interfacial shear and its link to the bubble velocity and curvature, and spectral properties of velocity and pressure, Figure 1,2, Eqs.(1-3). By diagnosing dependence of these quantities on the density ratios, flow symmetries, initial conditions, and accelerations, by identifying their universal properties, and by accurately measuring departures of data in real fluids from theoretical solutions in ideal fluids, one can further advance knowledge of RM dynamics in realistic environments, better understand RM relevant processes in nature and technology, and improve methods of numerical modeling and experimental diagnostics of RM dynamics in fluids, plasmas, materials.

To conclude, we have solved the long-standing problem of RMI with variable acceleration by applying group theory. We have directly linked the interface velocity, morphology and shear, revealed the interfacial and multi-scale character of RM dynamics, achieved excellent agreement with available observations, and elaborated new theory benchmarks for future experiments and simulations.

**Acknowledgements**

SIA thanks the University of Western Australia (AUS) and the National Science Foundation, (USA).

**Figure captions**

Figure 1: One parameter family of regular asymptotic solution for 3D flow with group p6mm at some Atwood numbers. (a) bubble velocity versus bubble curvature; (b) bubble velocity versus interfacial shear; (c) interfacial shear versus bubble curvature; (d) stability of regular asymptotic solutions.

Figure 2: Qualitative velocity field in laboratory reference frame for the Atwood bubble near the bubble tip (a) in the volume; (b) in the plane.

Figure 1

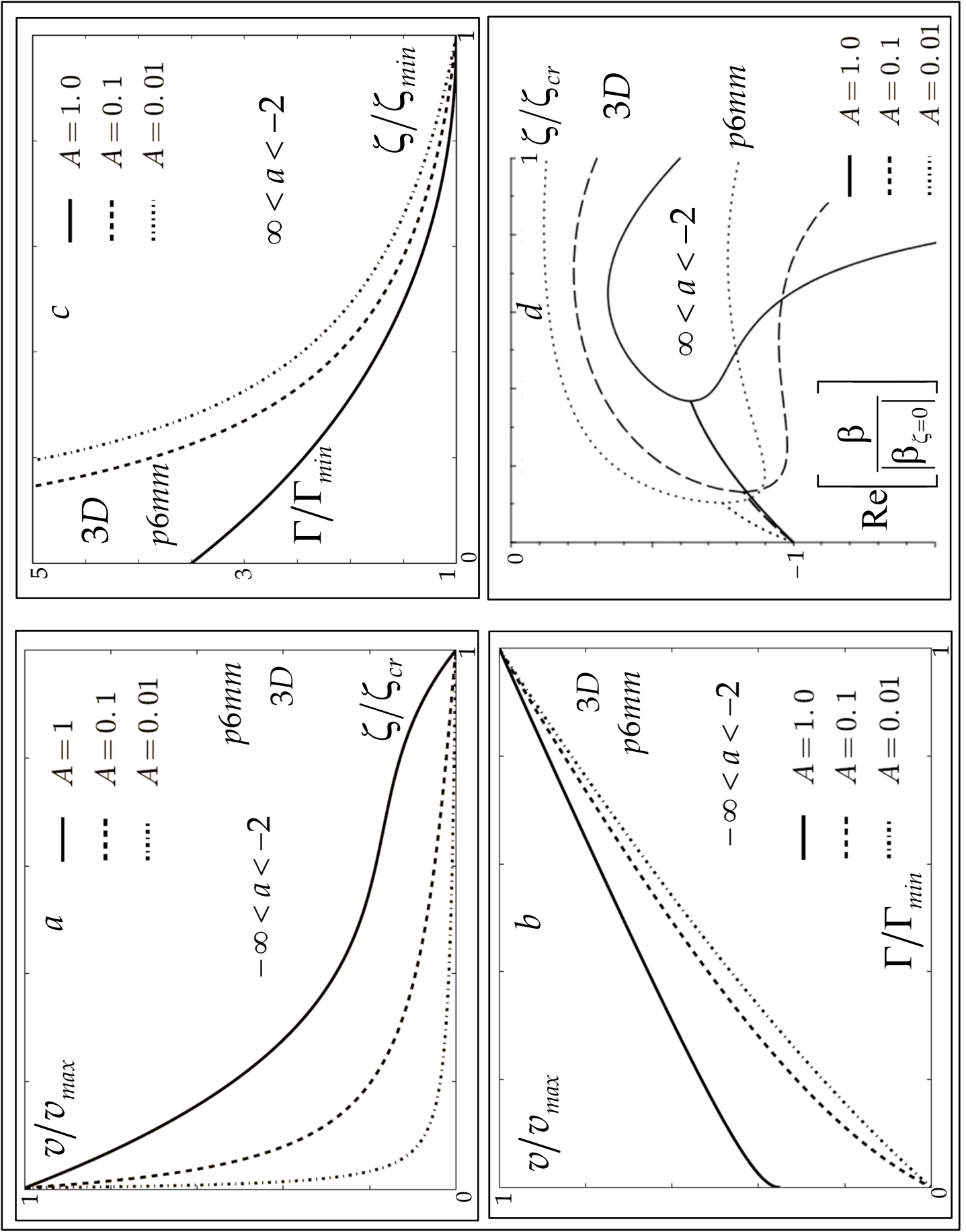

Figure 2

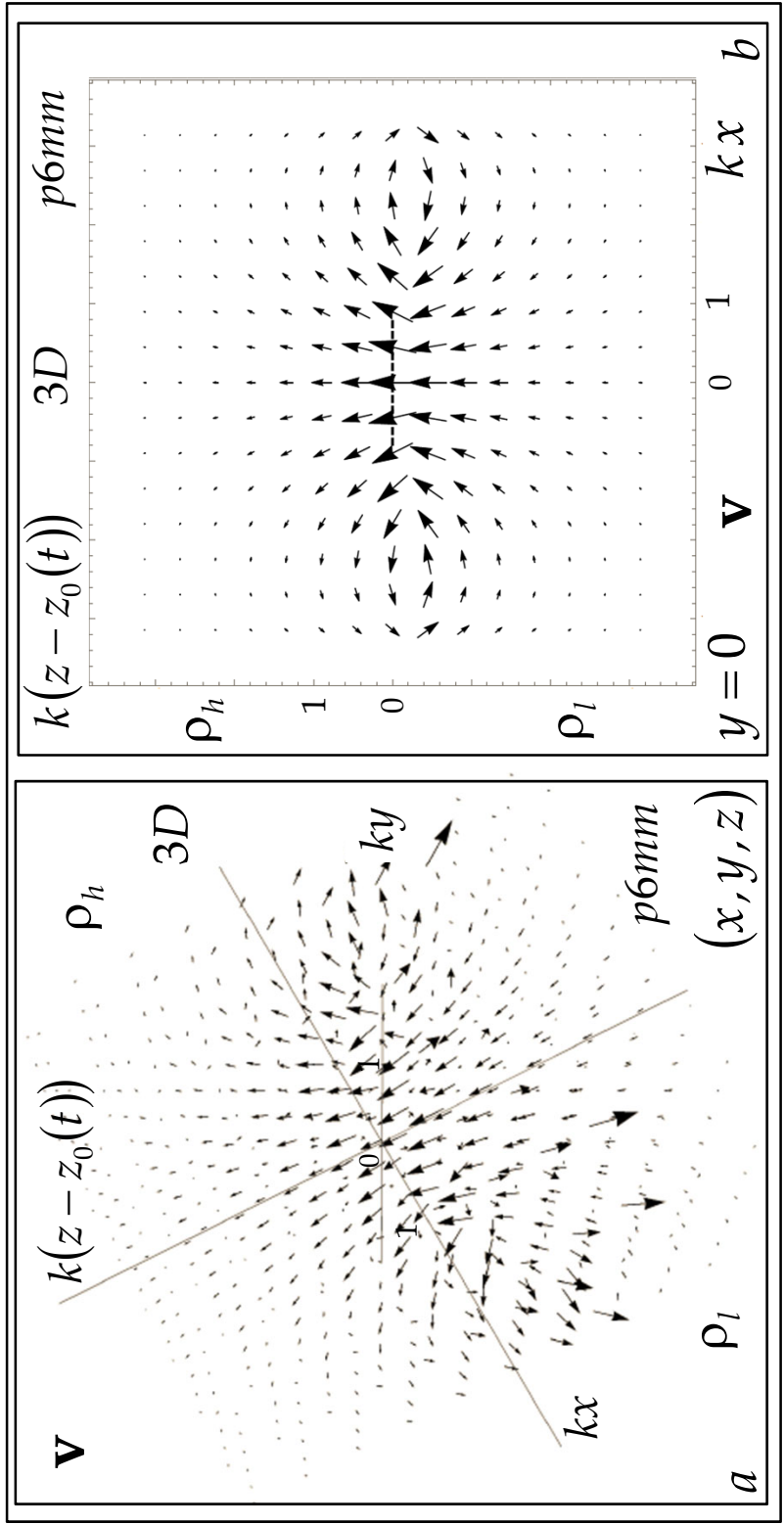